\documentclass[preprint]{aastex}

\usepackage{graphicx}
\usepackage{natbib}
\usepackage{color}
\usepackage{multirow}

\fontsize{6.5pt}{0pt}\selectfont

\slugcomment{Not to appear in Nonlearned J., 45.}

\shortauthors{Oba et al.}

\begin{document}
\title{Average radial structures of gas convection in the solar granulation}

\author{T. Oba\altaffilmark{1}}



\author{Y. Iida\altaffilmark{2}}


\author{T. Shimizu\altaffilmark{1}}

\altaffiltext{1}{Institute of Space and Astronautical Science, Japan Aerospace Exploration
Agency, 3-1-1 Yoshinodai, Chuo-ku, Sagamihara, Kanagawa 252-5210, Japan }
\altaffiltext{2}{Department of Science and Technology, Kwansei Gakuin University, Gakuen 2-1, Sanda, Hyogo 669-1337, Japan}

\begin{abstract}
\footnotesize{
Gas convection is observed in the solar photosphere as the granulation, i.e., having highly time-dependent cellular patterns, consisting of numerous bright cells called granules and dark surrounding-channels called intergranular lanes. 
Many efforts have been made to characterize the granulation, which may be used as an energy source for various types of dynamical phenomena. 
Although the horizontal gas flow dynamics in intergranular lanes may play a vital role, but they are poorly understood. 
This is because the Doppler signals can be obtained only at the solar limb, where the signals are severely degraded by a foreshortening effect. 
To reduce such a degradation, we use \textit{Hinode}'s spectroscopic data, which are free from a seeing-induced image degradation, and improve its image quality by correcting for straylight in the instruments. 
The dataset continuously covers from the solar disk to the limb, providing a multidirectional line-of-sight (LOS) diagnosis against the granulation. 
The obtained LOS flow-field variation across the disk indicates a horizontal flow speed of 1.8-2.4 km/s. 
We also derive the spatial distribution of the horizontal flow speed, which is 1.6 km/s in granules and 1.8 km/s in intergranular lanes, and where the maximum speed is inside intergranular lanes. 
This result newly suggests the following sequence of horizontal flow: 
A hot rising gas parcel is strongly accelerated from the granular center, even beyond the transition from the granules to the intergranular lanes, resulting in the fastest speed inside the intergranular lanes, and the gas may also experience decelerations in the intergranular lane. }
\end{abstract}

\section{Introduction}
\footnotesize{Solar granules are bright cells in the solar photosphere, surrounded by dark channels called intergranular lanes \citep{Stix2004}. 
A granule is a manifestation of an overshooting of a hot upflow from the convectively unstable subsurface layers into the stable photosphere \citep{Unsold1930}. 
The excess pressure above the upflow pushes the gas toward sideways. 
Radiative cooling, along with an associated increase in density, makes the material less buoyant, and the material begins to descend. 
A dark intergranular feature is formed at the location of the downflow. \\
　The enormous amount of the kinetic energy deposited in the granulation is transformed and may be used to drive various types of photospheric phenomena, e.g., \textit{exploding granules} \citep{Title1986, Rieutord2000, Hirzberger1999b, Fischer2017}, \textit{supersonic flows} \citep{Solanki1996, Nesis1993, BellotRubio2009, Quintero2016}, and \textit{reversed granulation} \citep{Rutten2003, Rutten2004, Leenaarts2005}. 
In addition, the granulation plays a role of an MHD wave driver, serving as the energy transfer in a non-thermal manner from the photosphere to the upper layers, the chromosphere and the corona (see \citealt{Parnell2012} for a review). 
Thus, the characterization of flow fields is remarkably beneficial for comprehending not only the granulation itself but also the other convection-driven phenomena. \\
　Our observational knowledge regarding the horizontal flow field still remains poor, compared to a vertical component of the flows, simply because its derivation is tricky.  
The Doppler signals for a horizontal component can be measured only near the solar limb \citep{Dravins1975, Beckers1978, Balthasar1985, Baran2010}. 
\citet{Beckers1970} first confirmed the observational existence of a horizontal outflow inside the granules, and detected the weaker amplitude of a horizontal flow (0.25 km/s) than that of an upflow (0.40 km/s). 
\citet{Keil1978}, \citet{Mattig1981}, and \citet{Nesis1989} found a twofold larger amplitude in the horizontal flows than that in the vertical ones. 
These authors reported an amplitude of a horizontal flow, which largely depends on the geometrical height, e.g., roughly 2.0 km/s, 0.7 km/s, and 0.6 km/s, respectively, at a geometrical height of approximately 100 km. 
\citet{ichimoto1989} and \citet{RuizCobo1996} attempted to depict the height structure of the vertical- and horizontal-flow fields, where the latter value is deduced from the vertical flow amplitude in accordance with the mass flux balance, i.e., div($\rho v$) = 0, where $\rho$ is the gas density and $v$ is the gas flow velocity. 
It should be emphasized that any solar limb observations are vulnerable to a spatial degradation owing to \textit{the foreshortening effect}. 
This effect causes an image degradation because the object's length along the LOS becomes shorter than the actual length as the observer's angle inclines toward the plane of the solar surface. 
Its spatial smearing effect is crucial even with the highest spatial resolution currently available. \\
　\textit{A feature tracking technique} has been developed and extensively used as an alternative approach over the last decades. 
This technique calculates the cross-correlation within two or more successive images, often observed at the center of the solar disk center, and measures the relative displacement of the moving features (mostly granules). 
The resulting speeds of the horizontal flows are found to be 0.8 - 1.3 km/s for the root mean square (RMS) \citep{Berger1998, Matsumoto2010, Keys2011, Manso_Sainz2011, Chitta2012}. 
By contrast, \citet{AsensioRamos2017} developed a totally different approach using a machine learning to find an immanent relation between the image intensity and horizontal gas flows in a numerically synthesized photosphere. 
It provides similar results to that of the feature tracking technique. \\
However, the feature tracking technique should be used with caution for a purpose of deriving the velocity fields at a spatial scale smaller than or even comparable to a granule, since this technique can characterize only distinctive shapes in an image (i.e., the rounded shape of a granule). 
Several authors investigated a limitation of the technique, by comparing the \textit{real} horizontal velocity fields in the granulation synthesized through a numerical simulation with the output retrieved by the technique: 
a poor agreement was particularly in intergranular lanes \citep{Louis2015} or at a small spatial scale ($<$1 [Mm]) \citep{Malherbe2018}, and the retrieved flow speed in RMS is smaller than the \textit{real} speed by a factor of 3 \citep{Verma2013} or the kinetic energy of a horizontal flow is underestimated by a factor of 5 to 6 \citep{Yelles2014}. \\
　To characterize the horizontal flow field at a small spatial scale down to a granule, we will return to Doppler analysis at the limb observation, by using the spectral data currently available with a good spatial resolution. 
The space-borne telescope onboard the \textit{Hinode} spacecraft \citep{Kosugi2007} is suited to meeting this requirement.
Its spatial resolution is the highest among those used in the previous studies that characterize the Doppler velocity fields over the center-to-limb. 
The space-borne telescope benefits from providing an almost constant data quality by eliminating time-dependent degradation such as an atmospheric seeing. 
Moreover, we attempted to minimize the image degradation owing to the instrumental stray light occurring in the \textit{Hinode}'s optical configuration, e.g., primary mirror with obscuration of the secondary mirror; 
the stray light degrades the image quality owing to the light property of diffraction, which distributes the incoming light over several pixels in the camera \citep{Schroeder2000}. 
\citet{Danilovic2008} investigated how much stray light in the \textit{Hinode}'s optical configuration degrades the image contrast of the granulation, and found that this degradation reduces the contrast by almost half. 
Recently, several authors have attempted to remove out the stray light, by applying \textit{deconvolution processing}, which restores the \textcolor{red}{real} image under a lack of the image degradation. 
The restored data shows an almost twofold larger contrast in intensity \citep{Mathew2009, Wedemeyer2009} and in the convective velocity fields \citep{Oba2017b}. 
Our study adopts a deconvolution technique in \textit{Hinode}'s spectral data, eliminating the major concern in a limb analysis. \\ 
　Our objective is to derive the amplitude of the horizontal flow and its spatial distribution in the solar granulation. 
The remainder of this paper is organized as follows. 
In Section 2, we describe the observations as well as our approach to calculating the horizontal flow fields. 
In Section 3, we report our observational results. 
In Section 4, we discuss the averaged nature of the flow structures in the granulation. 
Finally, in Section 5, we summarize our findings. \\

\section{Observations}
The observation data in this work were taken by a spectropolarimeter ($SP$; \citealt{Lites2013b}) used by the Solar Optical telescope (SOT) on board the \textit{Hinode} spacecraft \citep{Tsuneta2008,Suematsu2008,Shimizu2008,Ichimoto2008}. 
The \textit{Hinode}/SP observed quiet regions, recording the Stokes $I$, $Q$, $U$, and $V$ profiles of the Fe~{\sc i} 630.15 and 630.25~nm spectral lines with a spectral sampling of 21.5~m\AA. 
The spatial pixel sizes along the slit and scanning directions are approximately 0$\farcs$16 and 0$\farcs$15, respectively, with a spatial resolution of 0$\farcs$3. 
The integration time was 1.6 s per slit position. 
The slit orientation is fixed in the solar N--S direction. 
For the disk center observation, the observation was executed on August 25, 2009, between 08:01 and 09:59 UT. 
The FOV of $4\farcs5\times60\farcs9$ was repeatedly scanned 117 times with 30 slit positions per scan with an average cadence of 62~s. 
For the center-to-limb observation, the SOT team members, including the authors, conducted raster scans from the solar disk center to the north limb, and thus the data did not incur a Doppler shift induced by the solar rotation (Table \ref{tab:dataset}). 
Portions of the $\mu$ ranges are overlapped with respect to each dataset such that the wide coverage of $\mu$ from 0.11 to 0.96 is continuous. 
The center-to-limb data employed the same observational setup (e.g., pixel sampling and integration time) as the disk center data, although its slit-length is replaced by a longer one (81$\farcs$2). 
The observation durations of these datasets typically reach 1 hour, which is considerably longer than the typical lifetime of the granulation (6 min; \citealt{Hirzberger1999}) and the oscillatory motions (5 min; \citealt{Deubner1974}). 
The SP data were calibrated using the standard routine SP\_PREP in the Solar SoftWare package (\citealt{Lites2013a}); 
this procedure performs i) dark field correction, ii) flat-field correction, iv) the correction of curved spectral line, iv) the removal of periodic wavelength and spatial shifts in the period of the spacecraft, and vi) the calibration of intensity variation along the SP slit caused by the thermal deformation of the instrumental optics.
Finally, a systematic error of the absolute velocity (i.e., the offset of the absolute wavelength position) was estimated in the same manner as \cite{Oba2017} such that the line core of our average spectral profile, integrated broadly over space and time, matches that in a well-calibrated spectral catalogue \citep{Allende1998}, falling within a systematic error of 0.18 km/s. \\
　This study focuses on \textit{normal} convection in hydrodynamics, without disturbance by magnetic elements. 
The dataset covers a quiet region but also unavoidably covers magnetized atmosphere. 
To see how much the atmosphere is magnetized, we computed the polarization profile (Stokes $V$) averaged over the entire FOV in each dataset.
The degree of polarization is described by the peak values of the profiles, of which the highest value reaches 0.15$\%$ relative to the continuum intensity, and the other profiles range between 0.01$\%$ and 0.06$\%$. 
We conjecture that such a tiny contamination by a magnetized atmosphere should not significantly affect the statistical results. 

\section{Methodology}
\subsection{Spatial deconvolution}\label{bozomath}
To correct for the instrumental stray light contamination, the spectral data undergo a deconvolution processing (see \citealt{Jansson1997} for an overview). 
The processing used is the same algorithm as that developed in \cite{Oba2017b}, which is based on the Richardson-Lucy algorithm \citep{Richardson1972, Lucy1974}, incorporating a regularization term designated to suppress the noise amplification \citep{Dey2004}. 
This algorithm requires information of a point-spread-function (PSF) for the employed instrument, provided here by \cite{Danilovic2008}, who modeled the PSF including the major factors of image degradation induced by the optical configuration of the SOT. 

\subsection{How to derive the LOS flow field}
　We processed the deconvolved spectral profiles into velocity fields through a \textit{bisector analysis} \citep{Dravins1981}, the detailed setup of which is described in \cite{Oba2017, Oba2017b}. 
A bisector gives a wavelength position dividing the spectral line in two equal width, and its placement from the original wavelength is attributed to the Doppler shift at various intensity levels. 
Computing the bisectors of the Fe~{\sc i} 630.15~nm spectral line, we obtained the LOS velocity fields at six bisector levels, defined here by $I/I_{0}$=0.70, 0.65, 0.60, 0.55, 0.50, 0.45, where $I$ is intensity level at each spatial pixel, and $I_{0}$ is the continuum intensity averaged over \textit{the disk center}. 
Note that this $I_{0}$ definition is also adopted for the non-disk center observations, aiming to provide the same temperature layer sampling irrespective of the heliocentric angle. 
The physical meaning of such an approach is that any observed spectral profiles are normalized with the \textit{absolute} intensity, i.e., rather, the bisector levels are defined using the data number (DN) as the unit. 
Thus, our bisector analysis aims to sample the identical temperature layer even at a different heliocentric angle under the following rough approximation; 
any observed light with a certain intensity originates from the corresponding temperature layer through the Planck function, valid within the optical thick regime and in the first-order approximation of the temperature stratification along the optical depth, which is moderately sufficient for the solar photosphere \citep{Gray2008}. 
The stratification of the photospheric temperature stratification is nearly invariant with respect to the solar latitude, namely, as large as an enhancement of 2-3 K at the pole \citep{Rast2008}, corresponding to a geometrical height difference of a few km. 
Consequently, the bisector intensity levels sampled at the absolute DN reflect the same temperature layer,  irrespective of the heliocentric angle. 
Intensity levels of 0.70-0.45 are related to the Planck function temperature, through the average quiet Sun model of \citet{Vernazza1981}, then proving the corresponding geometrical heights of roughly from 50 km to 140 km.
The calculated geometrical height should be accepted when the temperature stratification holds the plane-parallel structure although the actual photosphere does not, as the corrugation of the iso-optical depth surface is estimated to 30 km/ in rms \citep{Stein1998}. 
The intensity range of $I/I_{0}$=0.70 to 0.45 is included between the line core and continuum even at an off disk-center, except at the extreme limb. 
Fig.\ref{fig:prof} confirmed this fact, showing spectral profiles normalized with the disk center continuum intensity at several different heliocentric angles. 
Spectral profiles separately averaged in granules ($I_{c}>$0) and intergranular lanes ($I_{c}<$0) are illustrated on the top and bottom panels, respectively, where $I_{c}$ is the average continuum intensity at each heliocentric angle.
One noticeable concern in our dataset, intermittently observed from 2007 to 2017, is an instrumental throughput degradation over time; \citet{Lites2013b} reported a decreasing trend of approximately 17 \% in five years. 
To minimize the throughput difference among our dataset, by referring to irradiance scan data regularly executed every month from the beginning of the \textit{Hinode} launch, we normalize the DN in each dataset to be identical with this reference data at the closest timing (Table\ref{tab:hop79}). 
Fig.\ref{fig:illus} demonstrates the validity of such correction, showing smooth variation of the continuum intensity across the center-to-limb.
Our elaborate bisector processing mentioned above provides a reliable sampling of the same temperature layer across the solar disk. \\

\subsection{Removal of the 5-min oscillation}
The derived velocity field includes another-origin (not the convection-origin), namely, the 5-min oscillation. 
This oscillation is an assembly of numerous numbers of eigenmodes generated with the solar spherical stratification \citep{Leighton1962}, and typically fluctuates 0.3-0.4 km/s in RMS in the photospheric velocity fields \citep{Ulrich1970}. 
A \textit{subsonic filter} \citep{Title1989, Rutten2004, Matsumoto2010} is a well-established technique for extracting the convective velocity alone from a time series of velocity map by filtering out the 5-min oscillations. 
Our study adopted the same setup as described in \citet{Oba2017, Oba2017b}. 
Each dataset from the disk center to the limb undergoes the above filtering processing separately. 
In all the created power spectra in the time-spatial frequency domain, the 5-min oscillatory power seems to be located in the identical frequency domain among the dataset, whereas the power decreases toward the limb presumably owing to the predominant vertical fluctuations \citep{Stix1974, Schmidt1999}. 

\subsection{Deriving the horizontal flow}
We introduce how to determine $v_{h, std}$ (horizontal flow speed in standard deviation) from the LOS amplitude variations over the disk-center to limb. 
Any LOS flow fields, except for when observed at the disk center, consist of components vertical and horizontal to the solar surface. 
We assume that the granulation has no latitude-dependence and that flows in the vertical and horizontal directions are uncorrelated. 
Under this assumption, one of the horizontal components $v_{x}$ (defined by the spectrograph slit direction projected on the solar surface plane) can be related to the other observable physical quantities through the following equation: 
\begin{equation}
v_{los, std}^2 (\mu)=\mu^{2} v_{z, std}^{2}+(1-\mu^{2}) v_{x, std}^{2},
\label{eq:vh}
\end{equation}
where $v_{los, std}$ is the observed convective velocity in the standard deviation along the LOS, $v_{z, std}$ stands for vertical motion obtained at the disk center, $v_{x, std}$ stands for the horizontal motions we aim to derive, and $\mu$ is the heliocentric angular distance represented conventionally as $\mu=\textrm{cos} \ \theta$; here, $\theta$ is a heliocentric angle. \\
　Past studies \citep{Keil1978, Mattig1981} have also applied the above equation but with the standard deviations replaced by the RMS. 
Their purpose is to derive $v_{x}$ averaged over granules and intergranular lanes, where the RMS is also valid as long as the average value of the vertical flow speed holds at zero. 
Our interest is to derive $v_{x}$ in the granules and intergranular lanes separately, which should be derived through the standard deviation form, because each average vertical flow is non-zero as the granules tend to have an upflow and the intergranular lanes have a downflow. \\
　The fitting process provides a reasonable value of $v_{x, std}$, which minimizes the dispersion of $v_{los, std}$ versus the resulting value of the right-side of the equation. 
Note that $v_{los, std}$ reflects only one of the two directions along the horizontal plane, although the gas flow along its perpendicular direction actually exists. 
In symmetric horizontal gas flows, the actual amplitude should be given through
\begin{equation}
v_{h, std}= \sqrt{2} v_{x, std}. 
\label{eq:hor}
\end{equation}

\section{Results}
\subsection{LOS velocity field}
First, we report the obtained the center-to-limb variation of $v_{los}$. 
Fig.\ref{fig:clv} shows the arbitrary chosen snapshots of the continuum intensity (top panels) and the co-spatial $v_{los}$ at a bisector level of 0.45 (bottom panels), at a different $\mu$ of 1.0, 0.8, 0.6, and 0.4 from the left to right panels. 
The contrast of $v_{los}$ is the smallest at $\mu$=1.0 and the largest at $\mu$=0.6, as shown in their RMS values of 0.79 and 1.09 km/s, respectively. 
As the heliocentric angular distance $\mu$ decreases (toward the limb), the foreshortening effect conspicuously appears: granular and intergranular shapes become compressed along the north-south direction while keeping their length along the east-west direction, resulting in elliptical shapes in their appearance. 
In the $v_{los}$ maps, we can find two characteristics along the south-north and east-west directions, respectively. \\
　Regarding the south-north direction, the peak of the blue-shifted signal does not coincide with the center of the granule; 
the front (south) side of the granules generally displays strong blue-shifted signals, whereas the rear (north) side generally shows weak blue-shifted or occasionally red-shifted, signals (Fig.\ref{fig:clv}). 
This trend is particularly seen in the slices sandwiched by the two arrows facing each other, indicated by \textit{a} and \textit{b}, which upon closer inspection are displayed in the left panels of Fig.\ref{fig:slice}. 
In the granules, the peak of the blue-shifted signal is offset toward the front side (i.e., the south side) from that of the continuum intensity, as shown in 0.1-0.9 Mm and 1.4-2.0 Mm of \textit{sample a}, and more clearly seen in 0.4-1.6 Mm of \textit{sample b}. 
This asymmetry between the south and north sides of the granule is in good agreement with a granular radial flow; 
when viewed from the inclined observer's angle, the south side reflects a horizontal flow approaching toward the observer (i.e., blue-shifted), while the north side shows a horizontal flow moving away from the observer (i.e., red-shifted). \\
　For the other characteristic, the existence of a non-radial flow is implied by the spatial relation between $v_{los}$ along the east-west direction and the granulation morphology. 
The dark intergranular lane, sandwiched by adjoining granules along the east-west direction, is filled with a blue-shifted signal. 
Conspicuous examples of such intergranular lanes are seen in the slices indicated by the arrows \textit{A} and \textit{B} in Fig.\ref{fig:clv}, the plots of which are shown in 0.4-1.3 Mm of \textit{sample A} and 1.5-2.0 Mm of \textit{sample B}. 
This trend can be explained by the granular non-radial flow, because $v_{los}$ on the west and east sides are not aligned in the radial direction. 
Our finding of radial flow is not surprising if the gas convection holds the mass conservation. 
The other finding does not come from a natural consequence of the gas convection, indicating a non-radial flow can develop strongly enough to be observed. \\
　By contrast, we cannot clearly see the opposite trends: 
namely, for a slice of intergranular lane cutting through the south-north direction, no clear trend of weak red-shifted signals at the front side or strong red-shifted signals at the rear side, and for a slice cutting through the east-west direction, no red-shifted signal associated with intergranular lanes sandwiched by adjoining granules. 
We attribute the lack of a clear indication of these opposite trends to still the remaining lack of spatial resolution because low-intensity signals in intergranular lanes are strongly diluted with higher-intensity signals from granules and are preferentially lost. \\

\subsection{Amplitude of horizontal flow field}
Fig.\ref{fig:clv_rms} describes $v_{los, std}$ variation over the center-to-limb at a bisector level of 0.55. 
In this Figure, we calculates all the standard deviation values along the slit direction with a constant spacing of 0$\farcs$16, and arranges the values in order of heliocentric angular distance. 
The standard deviation is calculated from pixels covering whole the repeatedly scanned duration for each dataset (typically more than 1 hour) and covering 2 Mm within the center part of the scanned range, since pixels close to the edge are uncertainly deconvolved as they need information outside the FOV. 
Each of the pixel coverages correspond to 2 and 10 samples of a typical extent and lifetime of a granule, respectively.
Fig.\ref{fig:clv_rms} represents an arc-shape variation of $v_{los, std}$ over the center-to-limb; 
the amplitude starts from 0.82 km/s at $\mu=$1.0 and monotonically increases toward the limb, reaching 1.22 at $\mu=$0.6, and decreases at $\mu<$0.6. 
This arc-shape variation is a net consequence of the two competing factors: 
The decreasing factor of $v_{los, std}$ toward the limb is due to the foreshortening effect, whereas the increasing factor is due to the faster horizontal flow than the vertical flow. 
Although this increasing factor affects at the limb ($\mu<$0.6), the decreasing factor there becomes effective enough to overcome the increasing trend. \\
　For a reasonable estimation of $v_{h, std}$, we have to make a proper selection of which heliocentric angles should be adopted into the fitting (Eq.\ref{eq:vh}). 
This fitting function should be adopted to samples at as many heliocentric angles as possible but simultaneously under the minimal foreshortening degradation. 
Accordingly, we adopted four different coverages of the heliocentric angles, i.e., $\mu$ = 0.6-1.0, 0.7-1.0, 0.8-1.0, and 0.9-1.0, one example of which is shown with the gray-solid curve by adapting $v_{x, std}$ =1.38 km/s (or  $v_{h, std}$=1.95 km/s) into the fitting. 
The reason for eliminating the remaining angles (i.e., $\mu<$0.6) is that Eq.\ref{eq:vh} does not model the decreasing factor owing to the foreshortening degradation. 
Table \ref{tab:obs_hor} summarizes the amplitudes resulting from the fitting with a combination of the six bisector levels and the four fitted ranges. 
When only heliocentric angles close to the disk center (e.g., $\mu$=0.9-1.0) are adopted to the fitting, the estimated $v_{x, std}$ becomes higher than those with a wide coverage of the heliocentric angles (e.g., $\mu$=0.6-1.0). 
This is due to a less prominent foreshortening degradation close to the disk center. 
Whereas different fitted ranges result in a different value of $v_{h, std}$, those at a bisector level of 0.70 are roughly 1.8-2.4 km/s, because the maximum value (2.43 km/s) is derived with a fitted range of $\mu$=0.9-1.0 and the minimum value (1.93 km/s) is derived with a fitted range of $\mu$ = 0.6-1.0. 
This dependence on the fitting range induces a complexity in strictly determining the value of $v_{h, std}$. 
Instead, we determine the value of $v_{h, std}$ by averaging over four outputs with four fitting ranges, which are listed in the right-most column in Table \ref{tab:obs_hor}. 
For greater simplicity, the horizontal flow speed, averaging over all the bisector levels from 0.45 to 0.70 is 2.10 km/s. \\
　We make a crude estimation of the error of $v_{hor, std}$ derived through the fitting. 
Referring to a standard deviation of 0.39 km/s in one of the value $v_{los, std}$ in Fig.\ref{fig:clv_rms}, 
we distributed the random fluctuation of 0.39 km/s on the $v_{los, std}$ variation over $\mu$=0.60-1.00 and searched for the best fitted $v_{hor, std}$ for a hundred of time, finding the standard error is 0.04 km/s. 

\subsection{Horizontal flow fields in granules and intergranular lanes}
　The standard deviation of $v_{los}$, when viewed from an inclined heliocentric angle, can be attributed to the amplitude of $v_{h}$ as its flowing direction is randomly distributed. 
Fig.\ref{fig:ex_fit} shows scatter plots between $I_{c}$ and $v_{los}$. 
The left panel ($\mu=$1.00) presents the relation of a granular upflow and intergranular downflow with a somewhat small scatter.
On the right panel ($\mu$=0.60), the aforementioned relation still holds, but with a relatively larger scatter. 
The degree of this scatter clearly depends on $I_{c}$ being relatively large in a dark region (intergranular lanes) and small in a bright region (granules). 
To determine $v_{h}$, we process the fitting approach in the same way as introduced in section 4.2, but separately at 
several continuum intensity regions segmented with a spacing of $I_{c}$=0.04. 
The horizontal flow speed is also determined by averaging over the four best-fitted values derived with the four heliocentric angular coverages. 
Fig.\ref{fig:h} provides the obtained height structure of the horizontal flow amplitudes, as well as the structure of the vertical flow for a comparison with the relation of the granular upflow and intergranular donwflow. 
The horizontal flow speeds are generally faster in the dark regions than those in the bright region because the maximum speed occurs inside the intergranular lanes, $I_{c}$=0.92-0.96, and the amplitude decreases monotonically toward either side of the bright or even darker region. 
With any choice of $\mu$ ranges of the fitting (i.e., $\mu$ = 0.6-1.0, 0.7-1.0, 0.8-1.0, and 0.9-1.0), the fastest speed is at the same location ($I_{c}$=0.92-96). 
Note that the range of $I_{c}=$0.76-1.40 displayed in Fig.\ref{fig:h} covers almost the entire region (98\%), whereas our bisector analysis cannot sample brighter granules ($I_{c}>$1.40) or darker intergranular lanes ($I_{c}<$0.76). 
One remark is that some of granules depart from the average sense, e.g., a faint granule as displaying upflow but with $I_{c}<$0, and accordingly Fig.\ref{fig:h} should be interpreted as it simply draws the \textit{average} structure. \\
　Table \ref{tab:graint} lists the horizontal flow amplitudes at each bisector level in the granular and intergranular regions, separately, each of which is derived from averaging over regions of $I_{c}>$1.00 or $I_{c}<$1.00, respectively. 
For more simplicity, the horizontal flow speed averaging over the six bisector levels is 1.58 km/s in the granules and 1.79 km/s in the intergranular lanes, i.e., the flow is faster in the intergranular lanes. \\

\section{Discussions}
\subsection{Horizontal flow amplitude}
　Our deconvolution provides a remarkably enhanced amplitudes in LOS velocity fields over the center-to-limb. 
Fig.\ref{fig:clv_rms} gives one example of $v_{los, std}$ as a function of the heliocentric angular distance, obtained at a bisector level of 0.55. 
This amplitude monotonically increases from 0.82 km/s at $\mu$ = 1.0 to 1.22 km/s at $\mu$ = 0.6. 
The estimated amplitude of $v_{h}$ should be 1.95 km/s to explain the mentioned center-to-limb variation ($\mu$=0.6-1.0). 
Note that the estimated value becomes large when considering the variation only at a higher $\mu$ range, e.g., the reasonable value for $\mu$=0.9-1.0 is 2.27 km/s. \\
　First, let us compare the obtained LOS amplitude variation with those of previous studies \citep{Keil1978, Mattig1981, Nesis1989}. 
Their highest amplitudes peak at $\mu$=0.6-0.8, whereas their sampling over a heliocentric angle is discrete, e.g., $\Delta \mu$ = 0.2, 0.4, and 0.4, respectively. 
Our maximum amplitude of 1.22 km/s at $\mu$=0.6 highly exceeds their maximum amplitude of 0.4-0.6 km/s. 
This fact implies that our spectral data after deconvolution processing provide resolved granulation even far from the disk center by mitigating the image degradation owing to the foreshortening effect. 
By contrast, their estimations of $v_{h}$ largely differ from each other, although their LOS amplitudes are comparable among them. 
\cite{Keil1978} estimated the horizontal flow speed of $\approx$ 2.0 km/s, compensating for several causes of the degradation other than the spatial smearing, e.g., an observed emergent intensity cumulatively originates from several layers along the LOS direction, making the observed signal smoothed over the physical information. 
With a better spatial resolution achieved using a balloon-borne 30-cm telescope whose data are almost free from an atmospheric seeing degradation \citep{Mehltretter1978}, \citet{Mattig1981} estimated a flow speed of $\approx$ 0.7 km/s in RMS, and later \citet{Nesis1989} concluded a typical value of $\approx$ 0.6 km/s; the above values are taken at the geometrical height corresponding to our bisector sampling levels. 
Thus, such a threefold difference among their estimation implies that the horizontal flow speed heavily depends on their approach. 
Our result, simply using the spatially well-resolved data, indicates confidently a large amplitude (e.g., 2.10 km/s), even without treating the other causes of degradation. 
If we appropriately correct for the other degradation, this amplitude will be even larger. 
This additional effort is beyond our current scope; rather, we insist here that the horizontal flow speed is at least 1.8-2.4 km/s in terms of the standard deviation, derived using only treatment for the spatial resolution. \\
　Next, we compare our horizontal flow amplitude with that derived through feature tracking techniques in the past, as summarized in Table \ref{obs_vel}. 
Our value is the highest at 2-4 times larger than some of their values. 
We emphasize that a Doppler diagnostic approach can reach a higher amplitude than that derived through a feature tracking approach. 
Whereas feature tracking techniques work well at a spatial scale of larger than a typical granule, the Doppler approach benefits from having the capability of access at a small-scale, such as intergranular lanes and inside the granules. 
This is demonstrated through present study using currently available data with the aid of deconvolution processing. 
Furthermore, by improving the spatial resolution achieved in the future, a Doppler diagnosis will directly mitigate the image degradation owing to the foreshortening effect, thereby extracting more horizontal flows at a small scale. \\ 

\subsection{Average radial structures of gas convection}
Our study succeeded in providing horizontal flows in granules and intergranular lanes separately. 
Past observational studies (\citealt{ichimoto1989} and \citealt{RuizCobo1996}) have also drawn the height structure of horizontal flow fields, although they indirectly estimated them by referring to vertical flow fields with the mass balance imposed, i.e., calculated in accordance with div($\rho v$) = 0, where $\rho$ is the gas density. 
Our approach is instead to derive horizontal flow fields by analyzing the center-to-limb variation at each continuum intensity in the granules and intergranular lanes. 
There may exist a concern regarding whether the continuum intensities at different heliocentric angles represent the identical location in the granulation. 
This concern does not deserve serious consideration, however. 
In this study, $I_{c}$ is normalized by averaging over the entire FOV in each dataset to correct for the monotonic decrease of the continuum intensity toward the limb, which is widely known as \textit{limb darkening} \citep{Stix2004}. 
To check the validation of this normalization approach, we draw histograms of the normalized continuum intensity in each dataset at the disk center to the most off-disk center accessed in our study, $\mu$ = 0.6. 
These histograms (not shown) do not differ from each other, indicating that a continuum intensity normalized within each dataset represents more or less an identical location. \\
　Fig.\ref{fig:cartoon} illustrates the average radial structure of the gas convection when viewed as a slice of a granule, summarizing simply our findings introduced in section 4.3. 
Here, we discuss the sequential phases (a) and (b); the former represents the boundary between the up- and downflows, and the latter represents the location where the horizontal flow has the fastest speed. 
The location of phase (a) is fairly compatible with $I_{c}$ = 1.00, which is commonly referred to as outline granules (e.g., \citealt{Hanslmeier2000, Bello_Gonzalez2010, Abramenko2012}). 
This consistency confirms that $I_{c}$=1.00 is a place where the material's flow alters its traveling direction from upward to downward \citep{Khomenko2010}. 
The positional relation between phases (a) and (b), i.e., at $I_{c}$=1.00 and 0.94, respectively, implies the following convectional aspect, which has yet to be argued. 
A flowing gas parcel, even after turning into a downflow, still experiences a pressure gradient force horizontally directed from a granule to an intergranular lane, having the fastest speed in the intergranular lanes; 
the parcel subsequently decelerates as it proceeds toward the vertices of the intergranular lane. 
The enhanced pressure gradient force directing outside the granule will be locally created by the strong stratified atmosphere in the granules, created by gas excessively accumulated in the granular center, which sweeps the upwelling gas into the horizontal directions \citep{Nordlund2009}. 
Thus, we identified where the fastest horizontal flow occurs slightly outside a granule or inside the intergranular lane. \\
　Finally, we quantitatively compare our observational results with a synthesized three-dimensional atmosphere calculated by the recent numerical simulation code, namely, the MURaM code \citep{Vogler2005}. 
Fig.\ref{fig:num} describes a height structure of the velocity fields in the numerical simulation in a manner analogous to the observational result (Fig.\ref{fig:h}). 
In Fig.\ref{fig:num}, $I_{c}$ is calculated through the radiative transfer code of SPINOR \citep{Frutiger2000}, whose detailed setup is described in \citealt{Oba2017b}. 
Note that both vertical and horizontal flows are \textit{real} values (not the Doppler velocity), referring directly to the physical quantities in the numerical atmosphere. 
The velocity field is sampled at geometrical heights of 50-140 km, which are identical to those heights to which our bisector analysis refers. 
The fastest speed of a horizontal flow is at $I_{c}$=0.92-0.96 irrespective of the bisector level, which is identical to our observational result. 
Therefore, the synthesized granulation is fairly consistent with our observational findings, indicating that a numerical simulation actually reproduces well horizontal gas flows in the granulation. \\

\subsection{Limitation to the approach}
　We should remark a validity of the several assumptions made to derive the average radial structure of gas flow, and discuss how they are related to our results. \\
　First, we estimate how much the derived amplitude of horizontal flow includes flows with a spatial scale larger than the granulation, namely \textit{supergranulation} \citep{Rincon2018}. 
Its horizontal flow speed is 0.3-0.4 km/s and vertical flow is much slow, 0.03 km/s, with a spatial extent of roughly 30 Mm. 
In analogous way to the error estimation of $v_{los,h}$ (section 4.2), by distributing the amplitude of 0.3 km/s randomly on the $v_{los, std}$ variation over $\mu$=0.6-1.0 (0.3 km/s is multiplied by a factor of $\sqrt{1-\mu^{2}}$ to be projected in the LOS direction), where 24 numbers of supergranules exist, we found the best fitted $v_{hor, std}$ with an overestimation of 0.15 km/s. 
Consequently, the horizontal flow speeds listed in this paper are mostly the granulation alone, while approximately 8 percent is contributed from the supergranulation. \\
　Second, we consider effects associated with the limb observations other than the foreshoretening effect. 
The optical corrugation (as noted in section 3.2) is particularly expected to appear as a consequence of a radiative transfer, which is generally difficult to be taken in account. 
This effect forces the observer  at an inclined angle to see the larger projected area of granules and the smaller projected area of intergranular lanes \citep{Balthasar1985}, and may bias the normalization of the continuum intensity. 
Such a difference of the normalization value leads to possible ambiguity about whether a location of $I_{c}$=0.94 at $\mu <$1.0 corresponds to intergranular lanes seen at the disk center ($\mu$=1.0). 
We may think that our dataset, whose heliocentric angle extends up to $\mu$=0.6, still do not significantly suffer from such effect. 
Fig.\ref{fig:ave_vhm60} provides $v_{los}$ averaged over each continuum intensity separately at $\mu$=0.6, showing that a crossing point between upflow or downflow, i.e., a boundary between granules and intergranular lanes, falls into $I_{c}$=0.98-1.02. 
It should be emphasized that we did not intentionally make the null-velocity correction. 
We may conclude that, although the optical corrugation effect leads to an ambiguity of the consistency in the continuum intensity sampling between the disk center and the off disk-center, the fast horizontal flow actually happens in the downflow regions, i.e., in intergranular lanes. \\
　Finally, it should be emphasized that Fig.\ref{fig:cartoon} indicates a \textit{radial structure} of gas convection, in which the arrows are to represent \textit{horizontal} flow that degenerates both the radial and non-radial components. 
The cartoon of Fig.\ref{fig:cartoon} should be interpreted to be compulsorily projected onto a two-dimensional plane for convenience, while the existence of some non-radial flows (i.e., the perpendicular direction to the paper surface) are actually indicated by Figs.\ref{fig:clv} and \ref{fig:slice}. 
Our future interest is to depict radial and non-radial components separately, providing the three-dimensional flow fields in the granulation. \\

\section{Summary}
We studied the solar photospheric convection, focusing on horizontal gas motions, based on spectral data observed by the \textit{Hinode} spacecraft. 
This instrument is well suited to our study; 
the \textit{Hinode}'s spacecrafts's seeing-free observation easily maintains nearly a constant image quality among all of the center-to-limb dataset, which is a key requirement in calculating the horizontal flow amplitude. 
Its spatial resolution (0$\farcs$3) is the highest among all the previous studies charactering the horizontal flow speed. 
Moreover, our deconvolution processing reasonably corrects for the instrumental degradation from stray light, aiming to address one of the issues in a limb analysis. 
This study succeeded in providing a spatial distribution of the horizontal flows in the granules and intergranular lanes separately. 
Our observational findings can be summarized as follows: 
\begin{itemize}
\item The LOS amplitude monotonically increases from 0.82 km/s at $\mu$ = 1.0, reaching the maximum value of 1.22 km/s at $\mu$ = 0.6, and decreases toward the farther limb. 
\item The horizontal flow speed averaging over the granules and intergranular lanes ranges from 1.8 to 2.4 km/s. 
\item The horizontal flow speed is slower in granules with an amplitude of 1.58 km/s and is faster in intergranular lanes with an amplitude of 1.79 km/s. 
The maximum speed is observed after entering the intergranular lanes. 
\item The two-dimensional LOS flow field at the off-center disk indicates the existene of divergent flows radially from a granular center, and other non-radially streaming flows. 
\end{itemize}
Intergranular lanes, previously considered as simply a downflowing region, are now revealed to have another aspect, i.e., a strong horizontal flowing region. 
How the horizontal flow acts on the fluxtubes deserves a further investigation because such tubes have been observed to divert rapidly from the center of a convection cell and advected slowly along the cell boundary \citep{Simon1988}. \\
　We would like to emphasize that a spectral data analysis with a higher spatial resolution can shed light on the intergranular convective dynamics, achieved using a future telescope or even a currently available instrument if somehow treating stray light in the instruments. 
This would be helpful in disentangling the relation between the granulation and the phenomena produced in the photosphere, and hopefully even in the upper atmosphere. \\

\acknowledgments
\textit{Hinode} is a Japanese mission developed and launched by ISAS/JAXA, collaborating with NAOJ as a domestic partner, NASA and STFC (UK) as international partners. Scientific operation of the \textit{Hinode} mission is conducted by the \textit{Hinode} science team organized at ISAS/JAXA. This team mainly consists of scientists from institutes in the partner countries. Support for the post-launch operation is provided by JAXA and NAOJ (Japan), STFC (U.K.), NASA, ESA, and NSC (Norway). 
We are grateful to the \textit{Hinode} team for obtaining a series of datasets well suited to this analysis. 
We also thank S. K. Solanki for providing helpful comments and T. L. Riethm{\"u}ller for instructions regarding how to treat  numerical simulation code.  
The present study was financially supported by the Advanced Research Course Program of SOKENDAI and by JSPS KAKENHI, Grant NO. JP16J07106 and 18H05878. 

\bibliographystyle{apj}
\bibliography{myrefs}
\clearpage


\begin{figure}
\begin{center}
\includegraphics[width=12cm]{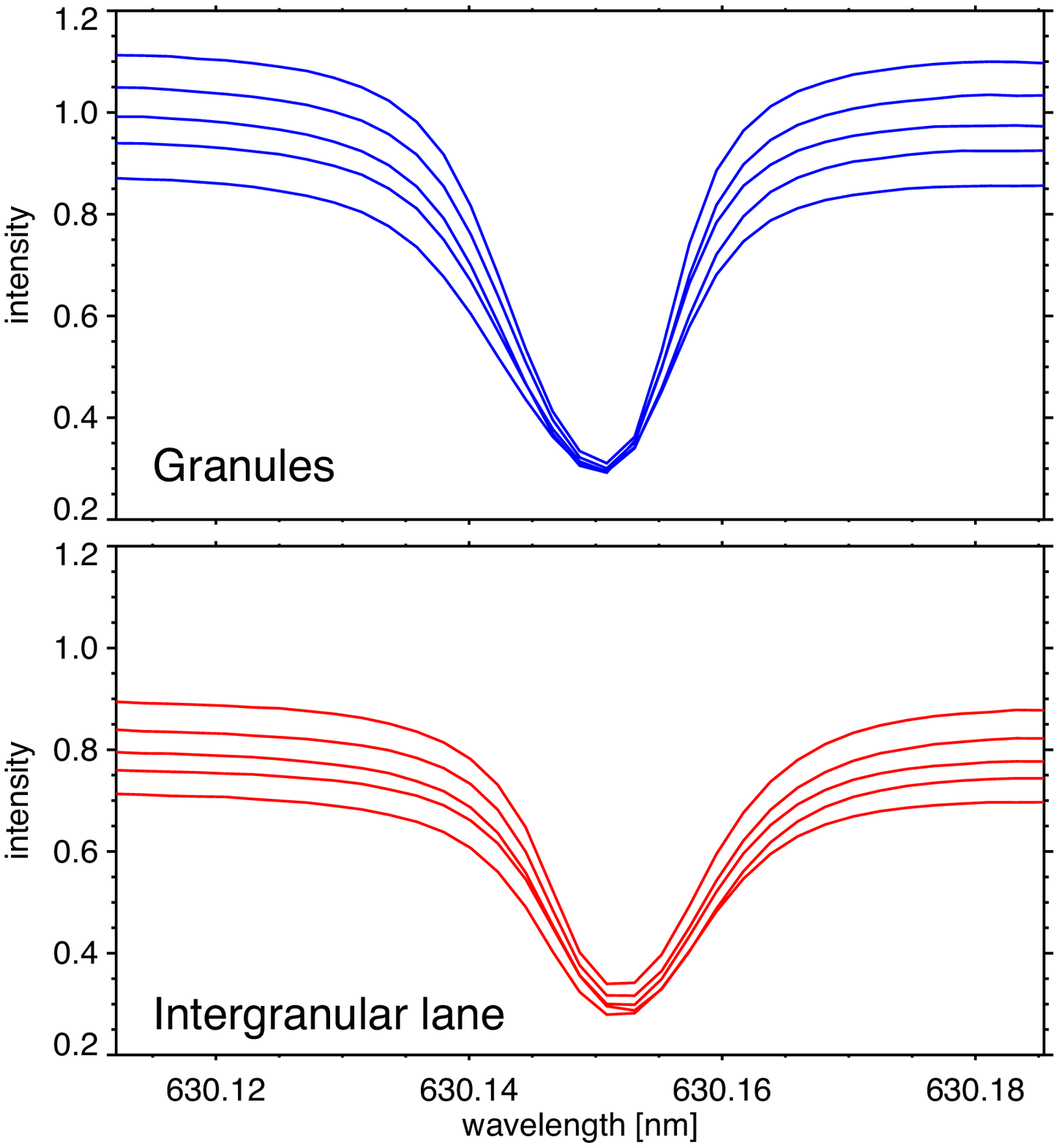}
\caption{Spectral profiles of Fe~{\sc i} 630.15~nm averaged over granules and intergranular lanes separately. Five spectral lines on each panel are at the different heliocentric angles of $\mu=1.00, 0.90, 0.80, 0.70, 0.60$, plotted from top to bottom. Note that all the spectral profiles are normalized at the continuum intensity at the disk center, thus the gradual decrease of the intensity toward the limb indicates the limb darkening effect. }
\label{fig:prof}
\end{center}
\end{figure}

\begin{figure}
\begin{center}
\includegraphics[width=12cm]{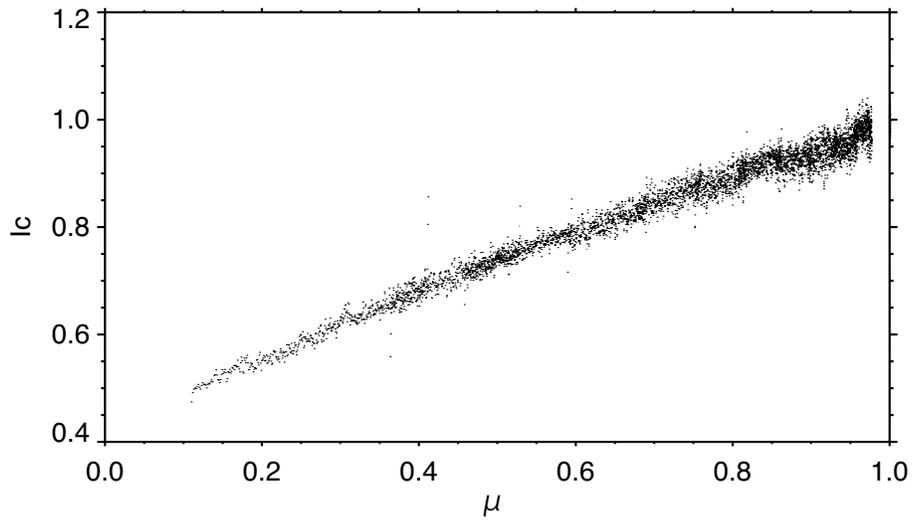}
\caption{Center-to-limb variation of the continuum intensity. 
This plot is created with all the pixels along the slit-direction (South-North) in a series of the dataset. 
Heliocentric coordinates of the pixels with a constant spacing of 0$\farcs$16 are transformed to heliocentric angular distance ($\mu$), which makes $\mu$ bin non-equidistant, as the dots are dense near the disk center in comparison with near the limb. 
Each dot is calculated from pixels that are in the center part of the scanned range over the repeatedly scanned duration for each dataset. }
\label{fig:illus}
\end{center}
\end{figure}

\begin{figure}
\begin{center}
\includegraphics[width=12cm]{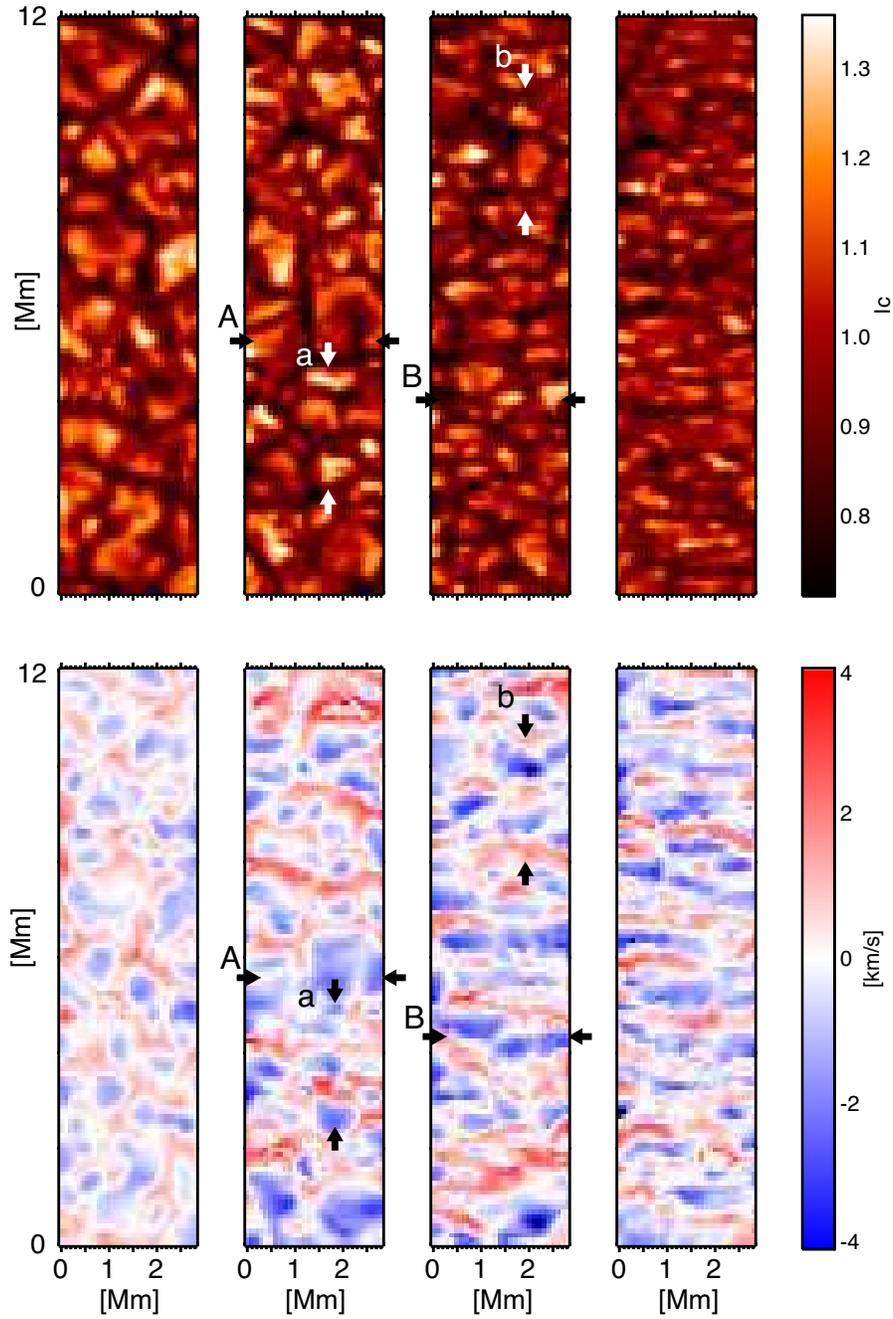}
\caption{Arbitrarily chosen snapshots of the continuum intensity (top panels) and the convective LOS velocity field at a bisector level of 0.45 (bottom panels), at the different angular distances of $\mu$ = 1.0, 0.8, 0.6, and 0.4, from the left to right panels. 
The plus signs of the x- and y-axes are oriented toward the west and north directions, respectively. 
The two arrows facing each other mark the sample slice, shown later in Fig.\ref{fig:slice}. 
Each of the four panels aligned at the top and bottom is described with an identical color contrast, following a single-color bar attached at the rightmost. }
\label{fig:clv}
\end{center}
\end{figure}

\begin{figure}
\begin{center}
\includegraphics[width=12cm]{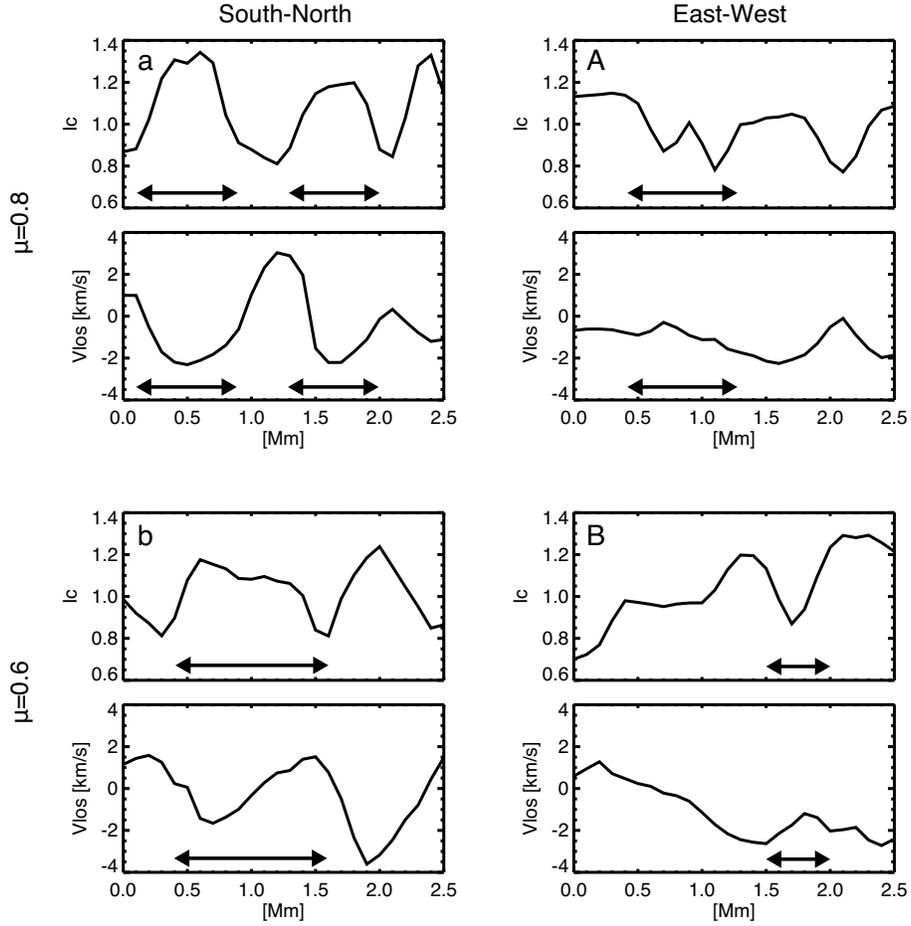}
\caption{The continuum intensity and convective LOS velocity fields on the slices, indicated by two arrows facing each other in Fig.\ref{fig:clv}. 
Those slices cut through south to north are plotted in \textit{samples a} and \textit{b} in the left panels, and through east to west are plotted in \textit{samples A} and \textit{B} in the right panels. 
The horizontal arrows provided at each bottom panel guide the eyes to identify the region of interest mentioned in the text. }
\label{fig:slice}
\end{center}
\end{figure}

\begin{figure}
\begin{center}
\includegraphics[width=12cm]{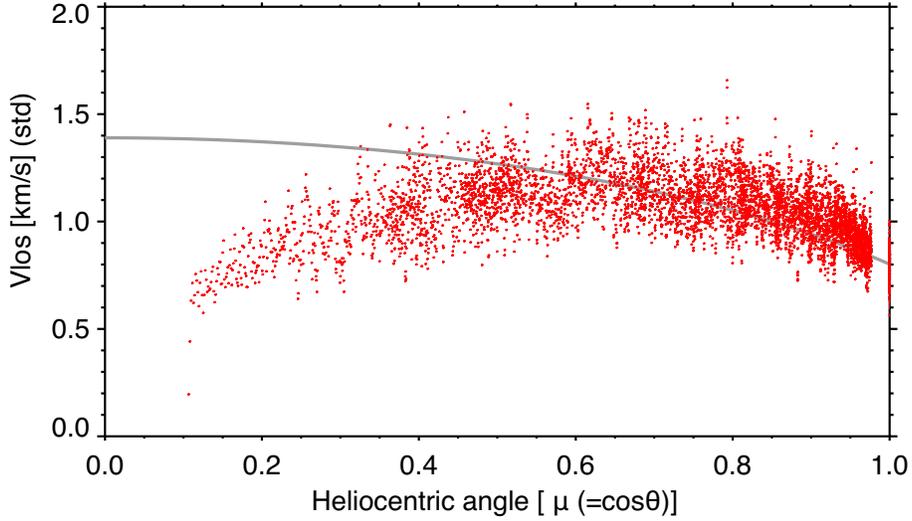}
\caption{Center-to-limb variation of the LOS velocity fields in the standard deviation at a bisector level of 0.55. 
The gray solid curve indicates the best-fitted result, adopting $v_{x, std}=1.38 $ km/s (or $v_{h, std}=1.95$ km/s) to the fitting equation. 
The LOS velocity fields at $\mu<$0.6 are discarded for the fitting such that the model fairly accounts for the center-to-limb variation over the rest of the angular distance. }
\label{fig:clv_rms}
\end{center}
\end{figure}

\begin{figure}
\begin{center}
\includegraphics[width=12cm]{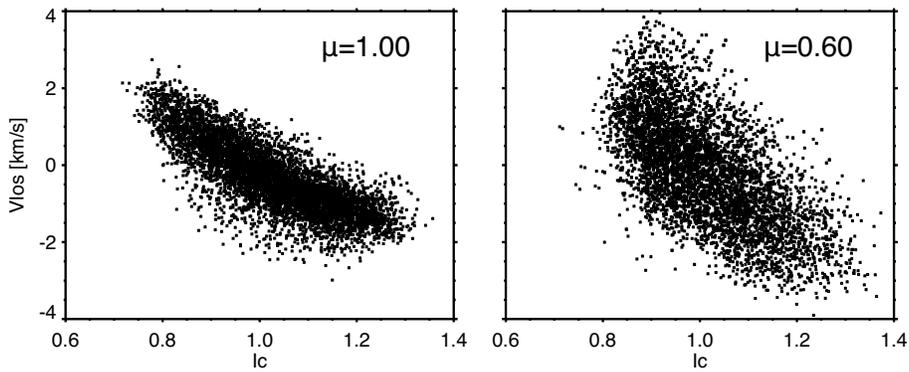}
\caption{Scatter plots of the continuum intensity versus LOS velocity field, at $\mu=$1.00 described in the left panel and at $\mu=$0.60 in the right panel. }
\label{fig:ex_fit}
\end{center}
\end{figure}


\begin{figure}
\begin{center}
\includegraphics[width=11cm]{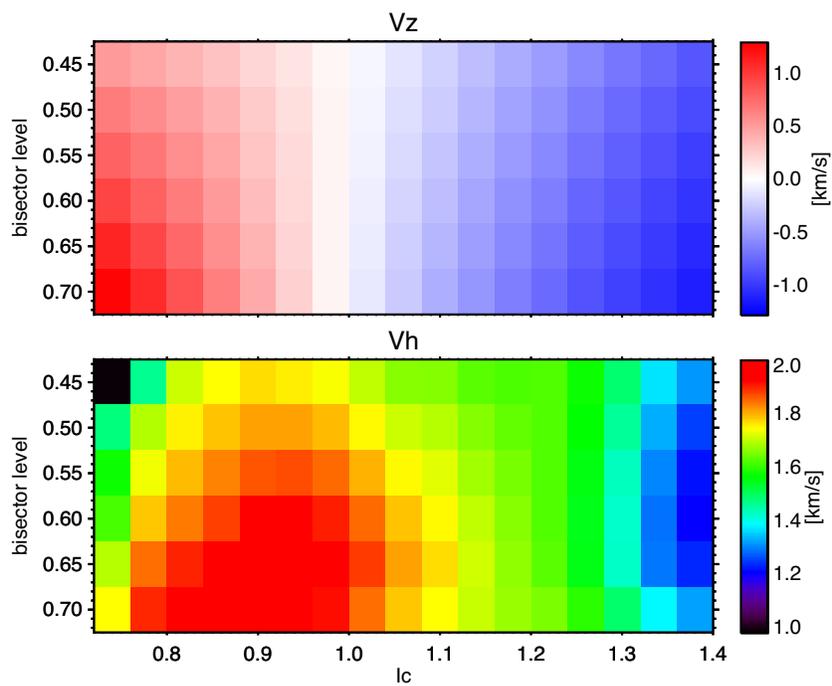}
\caption{Height structure of the vertical (top panel) and horizontal (bottom panel) flow speed. 
The former is calculated with the average value and the latter is calculated with the standard deviation, and both are derived after the deconvolution processing. }
\label{fig:h}
\end{center}
\end{figure}



\begin{figure}
\begin{center}
\includegraphics[width=12cm]{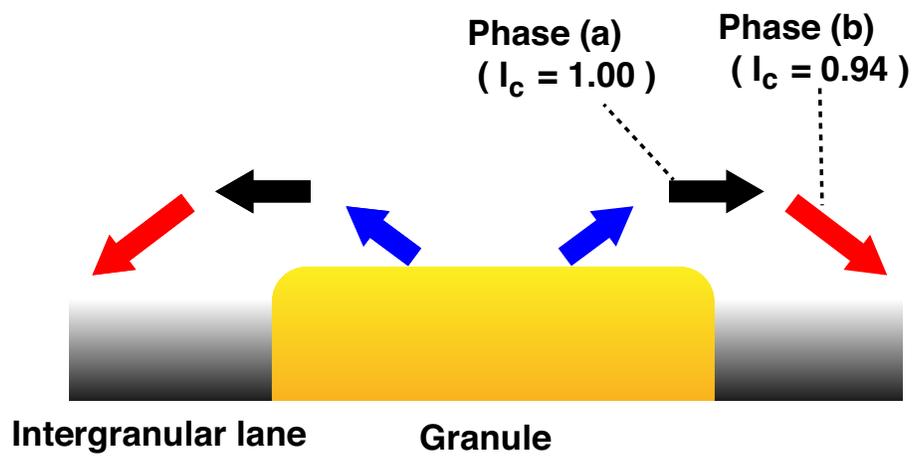}
\caption{Schematic slice view of vertical and horizontal flows in the granulation, where the arrow length indicates the relative flow speed, and the blue color identifies as an upward flow and the red color identifies a downward flow. 
Note that the arrows are to represent horizontal flow that degenerates both the radial and non-radial components. 
After passing through the transition from the granule to intergranular lane (phase:a), the horizontal flow speed reaches the fastest speed inside the intergranular lane (phase:b). }
\label{fig:cartoon}
\end{center}
\end{figure}


\begin{figure}
\begin{center}
\includegraphics[width=12cm]{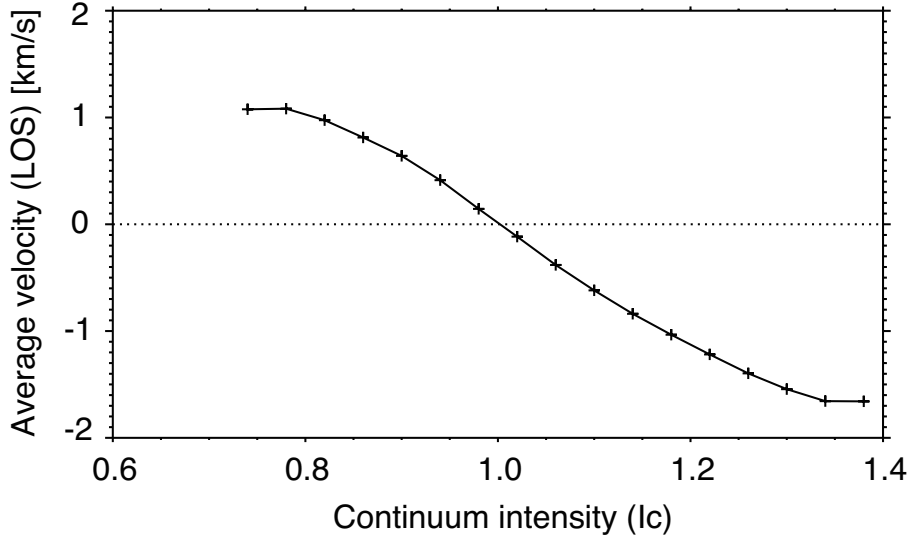}
\caption{LOS velocity fields at $\mu=0.60$, averaged at each continuum intensity separately. 
Note that a boundary between upflow and downflow regions falls into $I_{c}$=1.0, 
while a correction of the null-velocity has not been performed intentionally. }
\label{fig:ave_vhm60}
\end{center}
\end{figure}


\begin{figure}
\begin{center}
\includegraphics[width=11cm]{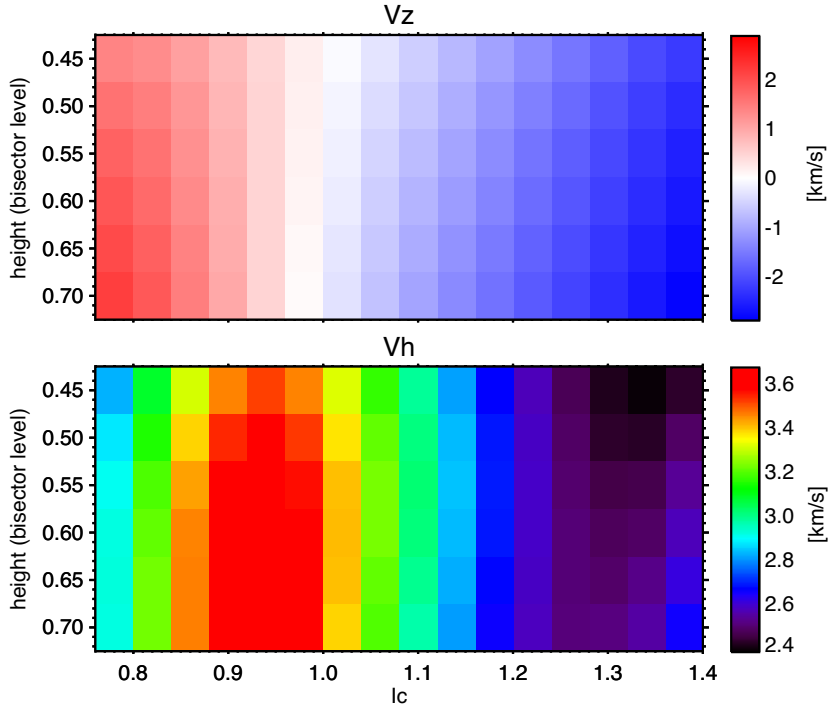}
\caption{
Height structure of a vertical (top panel) and horizontal one (bottom panel) flow speed in the reproduced atmosphere synthesized using the numerical simulation. 
The former is calculated using the average value, and the latter is calculated using the standard deviation in a manner analogous to the observational result (Fig.\ref{fig:h}). 
Note that the amplitudes directly refer to the physical quantities, not the Doppler velocity. }
\label{fig:num}
\end{center}
\end{figure}


\begin{table}
\begin{center}
\caption{Dataset description from the disk center to the limb \label{tbl-2}}
\begin{tabular}{ccccc} \hline
Dataset ID & Heliocentric angle (x, y) [arcsec] & $\mu$ & Date & Number of raster scan\\ \hline \hline
1 & (0, 0) & 1.00 & 2009/8/25 & 116\\
2 & (-9, 239) & 0.95-0.98 & 2017/7/07 & 58\\
2 & (0, 309) & 0.93-0.96 & 2017/5/19 & 58\\
3 & (0, 379) & 0.90-0.93 & 2017/5/17 & 48\\
4 & (0, 449) & 0.85-0.90 & 2017/5/13 & 71\\
5 & (1, 521) & 0.81-0.86 & 2017/4/28 & 98\\
6 & (4, 587) & 0.75-0.82 & 2017/4/13 & 68\\
7 & (0, 658) & 0.68-0.76 & 2017/4/12 & 66\\
8 & (6, 729) & 0.59-0.69 & 2017/4/08 & 69\\
9 & (23, 799) & 0.46-0.59 & 2016/8/23 & 141\\
10 & (0, 859) & 0.36-0.53 & 2017/2/19 & 53\\
11 & (2,920) & 0.11-0.41 & 2017/2/22 & 64\\ \hline
\label{tab:dataset}
\end{tabular}
\end{center}
\end{table}

\begin{table}
\begin{center}
\caption{Irradiance scan dataset referenced to correct for the time-degradation of the instrumental throughput. 
Reference ID listed in the right column denotes that our dataset in Table \ref{tab:dataset} is the closest in timing to the given irradiance dataset. }
\begin{tabular}{ccccc} \hline
Date & Reference ID in Table \ref{tab:dataset} \\ \hline \hline
2009/09/01 & 1 \\
2016/08/25 & 9 \\
2017/02/19 & 10, 11 \\
2017/04/14 & 5, 6, 7, 8 \\
2017/05/18 & 2, 3, 4 \\ \hline
\label{tab:hop79}
\end{tabular}
\end{center}
\end{table}

\begin{table}
\begin{center}
\caption{Horizontal flow amplitudes ($v_{h, std}$ [km/s]) with a combination of the six bisector levels and fitted ranges}
\begin{tabular}{c|cccc|c} \hline \hline
\multirow{2}{*}{bisector level ($I/I_{0}$)} & \multicolumn{5}{c}{angular distances ($\mu$)}\\ \cline{2-6} 
 & 0.6-1.0 & 0.7-1.0 & 0.8-1.0 & 0.9-1.0 & average \\ \hline \hline
0.45 & 1.84 & 1.90 & 2.02 & 2.08 & 1.96\\
0.50 & 1.90 & 1.97 & 2.10 & 2.18 & 2.03\\
0.55 & 1.95 & 2.02 & 2.16 & 2.27 & 2.10\\
0.60 & 2.00 & 2.07 & 2.22 & 2.35 & 2.16\\
0.65 & 2.02 & 2.10 & 2.26 & 2.41 & 2.19\\
0.70 & 1.93 & 2.06 & 2.23 & 2.43 & 2.09\\ \hline
average & 1.94 & 2.02 & 2.16 & 2.29 & 2.10\\ \hline \hline
\end{tabular}
\label{tab:obs_hor}
\end{center}
\end{table}

\begin{table}
\begin{center}
\caption{Horizontal flow amplitudes ($v_{h, std}$ [km/s]) in granular and intergranular regions}
\begin{tabular}{c|cccc} \hline \hline
bisector level ($I/I_{0}$) & granule (1.00$<I_{c}<$1.40) & intergranular lane (0.76$<I_{c}<$1.00 ) \\ \hline \hline
0.45 & 1.64 & 1.71 \\
0.50 & 1.66 & 1.78 \\
0.55 & 1.69 & 1.83 \\
0.60 & 1.71 & 1.88 \\
0.65 & 1.73 & 1.94 \\
0.70 & 1.72 & 1.95  \\ \hline
average & 1.58 & 1.79\\ \hline \hline
\end{tabular}
\label{tab:graint}
\end{center}
\end{table}

\begin{table}
\begin{center}
\caption{Reported amplitudes of horizontal flows, derived using a feature tracking technique \label{tbl-2}}
\begin{tabular}{lr} \hline \hline
　& Reported amplitude\\ \hline \hline
\citet{November1988} & 0.5-1.0 km/s\\ 
\citet{Brandt1988} & 0.67 km/s\\
\citet{Title1989} & 0.37 km/s\\
\citet{Berger1998} & 1.1 km/s\\
\citet{Shine2000} & 0.49 km/s\\
\citet{Muller2001} & 0.6 km/s\\
\citet{Matsumoto2010} & 1.1 km/s\\
\citet{Verma2011} & 0.54 km/s\\
\citet{Chitta2012} & 1.3 km/s\\ \hline
\end{tabular}
\label{obs_vel}
\end{center}
\end{table}


}

\end{document}